\documentclass[aip,preprint]{revtex4-2}
\usepackage{tikz}
\usepackage{pgfplots}
\usepackage{pgfplotstable}
\pgfplotsset{width=10cm,compat=1.9}
\usepackage[cp1251]{inputenc}
\usepackage[T2A]{fontenc}
\usepackage[english]{babel}
\usepackage{amssymb,latexsym,amsmath,amscd}
\usepackage{graphicx,color,framed}
\usepackage{mathrsfs}
\usepackage{hyperref}
\usepackage{bm}
\usepackage{appendix}
\usepackage{tabu}
\usepackage{tikz}
\usepackage{xcolor}
\usepackage{siunitx}
\usepackage{empheq}
\usepackage{float}
\usepackage{subcaption}
\usepackage{xr}
\usepackage{siunitx}
\usepackage{microtype}
\usepackage{amsfonts}
\usepackage{gensymb}
\usepackage[normalem]{ulem}
\graphicspath{{figures/}}

\makeatletter
\newcommand*{\addFileDependency}[1]{
  \typeout{(#1)}
  \@addtofilelist{#1}
  \IfFileExists{#1}{}{\typeout{No file #1.}}
}
\makeatother

\usepackage{color}

\begin{document}

\author{\firstname{Petr E.} \surname{Brandyshev}}
\email[]{pbrandyshev@hse.ru}
\affiliation{Laboratory of Computational Physics, HSE University, Tallinskaya st. 34, 123458 Moscow, Russia}
\affiliation{School of Applied Mathematics, HSE University, Tallinskaya st. 34, 123458 Moscow, Russia}
\author{\firstname{Yury A.} \surname{Budkov}}
\email[Author to whom correspondence should be addressed:]{ybudkov@hse.ru}
\affiliation{Laboratory of Computational Physics, HSE University, Tallinskaya st. 34, 123458 Moscow, Russia}
\affiliation{School of Applied Mathematics, HSE University, Tallinskaya st. 34, 123458 Moscow, Russia}
\affiliation{Laboratory of Multiscale Modeling of Molecular Systems, G.A. Krestov Institute of Solution Chemistry of the Russian Academy of Sciences, 153045, Akademicheskaya st. 1, Ivanovo, Russia}
\title{Theory of Casimir Forces: A Unified Approach Using Finite-Temperature Field Theory}
\begin{abstract}
We present a quantum theory of Casimir forces between perfect electrical conductors, based on quantum electrodynamics and quantum statistical physics. This theory utilizes Kapusta's finite-temperature field theory, combined with the Faddeev-Popov ghost formalism. This approach allows us to calculate Casimir forces at finite temperatures, providing both previously known and new physical insights from a unified perspective. Furthermore, our method enables us to compute the stress tensor associated with Casimir forces, in accordance with the Helmholtz free energy of an equilibrium quantum electromagnetic field. Using this method, we calculate the tangential pressure in a slit-like pore due to the Casimir effect.
\end{abstract}

\maketitle

\section{Introduction}

The Casimir effect is a fascinating phenomenon that occurs when quantum vacuum fluctuations of the electromagnetic field interact with conductive boundaries in space and time. This macroscopic effect, rooted in the principles of quantum mechanics and quantum field theory, has been extensively studied both theoretically and experimentally~\cite{bordag2001new,mostepanenko2009experiment,gong2020recent}. Through empirical evidence, scientists have been able to observe and quantify the Casimir force, providing valuable insights into the complex relationship between quantum field theory and the physical world~\cite{chan2001quantum,palasantzas2020applications,elsaka2024casimir}.

Since its initial discovery by Casimir~\cite{casimir1948attraction}, numerous research papers have been devoted to the development of mathematical techniques in quantum electrodynamics and quantum field theory to explain Casimir forces~\cite{barash1975electromagnetic,dzyaloshinskii1961general,schwinger1978casimir}. Great efforts were also made to more profoundly understand the physical nature of the Casimir effect~\cite{hoye1998van,hoye2017casimir,nikolic2016proof,nikolic2017zero}. Interestingly, it has been found that Casimir forces, arising from fluctuations, are not limited to quantum electrodynamics but also appear in various condensed phases such as liquid crystals, Coulomb fluids, critical fluids, {\sl etc.}~\cite{kardar1999friction,hertlein2008direct,ziherl2000pseudo,podgornik2001casimir,jancovici2004screening,ninham1997ion,hoye2017casimir,hatlo2008role,budkov2023variational}. The latter are also known as pseudo-Casimir forces~\cite{dean2009non,podgornik2001casimir} and critical Casimir forces~\cite{wang2024nanoalignment}. In these cases, the forces are driven by thermal or critical fluctuations of order parameters that characterize the state of the condensed medium, rather than quantum fluctuations of the electromagnetic field between metals or dielectrics in vacuum.

In all the aforementioned scenarios, the most natural mathematical framework for describing Casimir (or pseudo-Casimir or critical Casimir) interactions between macroscopic bodies involves field-theoretical methods, utilizing functional integration over fluctuations of order parameters, whether classical or quantum. When examining fluctuation forces in classical condensed matter, this approach is relatively straightforward, requiring the representation of a system's partition function through a functional integral over one or more fluctuating order parameters (scalar~\cite{budkov2023variational}, vector~\cite{blossey2022field}, or tensor~\cite{podgornik2021theory}) using the standard Hubbard-Stratonovich transformation. By employing the contact theorem~\cite{jancovici2004screening,dean2003field} or directly deriving the stress tensor from the thermodynamic potential of the field system~\cite{budkov2023variational,brandyshev2023noether,budkov2022modified,budkov2023macroscopic,golestanian1998path}, we can calculate the mechanical forces between macroscopic bodies, immersed in fluctuating medium. However, when tackling quantum systems such as the electromagnetic field~\cite{dudal2020casimir} or more sophisticated non-Abelian Yang-Mills gauge fields~\cite{chernodub2023boundary}, the issue of gauge ambiguity in quantum fields must be carefully considered. This ambiguity arises due to the presence of additional functional constraints on field variables, necessitating their inclusion in functional integrals. To address this challenge, finite-temperature quantum field theory, as developed by Kapusta and coworkers~\cite{kapusta2007finite}, provides a functional integration-based framework for quantum statistical physics that enables the systematic calculation of thermodynamic functions for gauge fields and matter fields in thermodynamic equilibrium. One way to resolve the problem of gauge ambiguity is by introducing auxiliary anticommuting (Grassmann) scalar fields, referred to as Faddeev-Popov ghosts~\cite{faddeev2010faddeev}. These fields help eliminate the complexities associated with gauge ambiguity, allowing for a more streamlined and rigorous calculation of thermodynamic functions in equilibrium quantum field systems~\cite{kapusta2007finite}. {However, for the quantum field systems in the thermodynamic limit, which have been only considered by Kapusta et al., there is no need to consider the problem of boundary conditions for ghost fields. In other words, there is a challenge in dealing with Faddeev-Popov ghosts in quantum field theories under confinement, as it is necessary to ensure consistency between the boundary conditions for ghost fields and the electromagnetic field at finite temperature. An example of such a system is the electromagnetic field in a nanosized pore with ideally conductive walls.}

In this paper, we will present a consistent theory of Casimir forces between perfect electric conductors, based on first principles of quantum electrodynamics and quantum statistical physics. The theory is derived from the Kapusta's finite-temperature quantum field theory supported by the Faddeev-Popov ghosts formalism. Note that this approach differs from previous field-theoretical models~\cite{bordag1985quantum,dudal2020casimir,canfora2022casimir,oosthuyse2023interplay,rahi2009scattering} that focused on calculating Casimir forces at zero temperature based on the boundary quantum field theory and scattering formalism. The present theory will allow us to calculate the Casimir force at finite temperatures reproducing the previously obtained and new physical results. Another strength of our approach is our method for calculating the stress tensor associated with Casimir forces, which is consistent with the free energy of the equilibrium quantum electromagnetic field. This will allow us to obtain stress tensors that satisfy the requirement of gauge invariance, thanks to additional terms introduced into the full stress tensor associated with the propagator of the ghost fields. As we will demonstrate below, these terms automatically exclude contributions from non-physical degrees of freedom of the electromagnetic field in the normal and tangential Casimir stresses. We will use this method to calculate the excess surface tension on the conductor surfaces due to the Casimir effect.

\section{Functional integral with boundary conditions}\label{PF}
Before we dive into the details of the method, let us discuss the basic principles behind it. We consider quantum field theory at finite temperature, which is formulated in the imaginary time. The action of this theory is invariant under ISO(4) transformations, in contrast to the action in standard quantum field theory, which, as is well known, is invariant under the Poincare group ISO(3,1). Thus, using Noether's first theorem, we can obtain the conserved tensor currents. In this case, we are talking about tensor Noether's currents associated with spatial symmetries. For example, we can obtain the stress tensor. In this case, the mechanical equilibrium condition acts as the conservation law~\cite{brandyshev2023noether,brandyshev2023statistical,budkov2022modified}. The gauge theory contains extra degrees of freedom due to gauge invariance. Since we consider the theory in the functional integral formalism, we use the Faddeev--Popov method to eliminate the extra degrees of freedom (see, for example, references~\cite{faddeev2010faddeev,kapusta2007finite}).

In order to apply Noether's first theorem to the effective action obtained as a result of this procedure, we must choose an explicitly covariant gauge-fixing condition. This is because the terms associated with gauge fixing that are introduced into the effective action should, as we assume, preserve the spatial symmetries of the original action. For example, we choose the covariant gauge condition $\partial_a A_a=0$, which is invariant under ISO(4) group. Then, Noether's first theorem can be applied to obtain currents related to both translation and rotational symmetries. We limit ourselves to considering translation symmetries, since we are only interested in the stress tensor. However, in general, our formulation of the theory allows us to consider rotational symmetries as well~\cite{budkov2024thermomechanical}, although this is beyond the scope of this paper. Thereby, we consider a theory with boundary conditions. In order to find the average value of the stress tensor, it is necessary to calculate the Green's functions for all fields, including ghost fields. Ghosts are needed to cancel out any extra degrees of freedom that would otherwise be present in the system. The Green's functions can be uniquely determined only if the boundary conditions they satisfy are known. The boundary conditions are initially imposed on the gauge field strengths, not on the gauge potential, which is reasonable since the strengths are observable. However, a well-known problem arises from the fact that integration in the functional integral is performed over the gauge fields, not their strengths. Therefore, it becomes necessary to impose boundary conditions on the vector potentials, and this may conflict with the requirements of gauge invariance of the original theory. Thus, for the boundary conditions to be invariant under gauge transformations $A'_a=A_a-\partial_a \theta$, it is necessary for the local gauge transformation parameter $\theta(x)$ itself to satisfy certain types of boundary conditions. Thus, the boundary conditions that the gauge transformation parameter satisfies follow from the requirement of gauge invariance of the boundary conditions for the vector potentials. Therefore, we show that the ghost fields must satisfy the same boundary conditions as the gauge transformation parameter. Therefore, the boundary conditions for the ghost fields are determined by the boundary conditions satisfied by the gauge fields. From this, we can obtain an explicit form of the Green's function for ghosts and substitute this function into the average stress tensor (\ref{stress}) to cancel out the contribution of extra degrees of freedom.

In addition, it should be noted here that the boundary conditions are formulated so that the fields satisfy the equations of motion (Maxwell's equations) at the boundary, which leads to additional boundary conditions for vector fields. This allows us to uniquely determine Green's functions for all fields. The equations of motion are satisfied only at the boundary because, in the interior points of the domain, the values of the fields are simply integration variables in the functional integral and do not satisfy any equations of motion. The fields are expanded in terms of eigenfunctions of an operator whose integral kernel is the kernel of a Gaussian measure. These eigenfunctions satisfy the boundary conditions imposed on the fields. The functional determinant of this operator is simply equal to the product of the eigenvalues corresponding to these eigenfunctions, allowing us to define the functional integral.

Thus, taking into account the above general considerations, let us move on to the technical details. Let us consider two perfectly conductive walls that are placed parallel to each other at a distance $L$ from one another (slit-like pore). Let the $x_3$ axis be perpendicular to the surfaces of the walls and the walls themselves be located at the points with the coordinates $x_3=0$ and $x_3=L$. The boundary conditions for the magnetic and electric fields are as follows:
\begin{equation}\label{1.27}
\begin{aligned}
\mathbf{E}_{||}(x_{1},x_{2},0,\tau)=\mathbf{E}_{||}(x_{1},x_{2},L,\tau) = 0,\\ \mathbf{B}_{\bot} (x_{1},x_{2},0,\tau)=\mathbf{B}_{\bot} (x_{1},x_{2},L,\tau) = 0,
\end{aligned}
\end{equation}
where we have defined
\begin{equation}\label{}
\mathbf{E}_{||}= (E_{1},E_{2},0),\quad \mathbf{B}_{\bot}=(0,0,B_{3}).
\end{equation}
Moreover, due to the absence of sources, we assume that at $x_{3}=0$ and $x_{3}=L$ the following equality is satisfied \footnote{Let us not forget that after the transition $t\rightarrow-i\tau$, for convenience we have redefined the electric field strength $\mathbf{E}\rightarrow i\mathbf{E}$.}
\begin{equation}\label{1.28}
\nabla\cdot \mathbf{E}=0,
\end{equation}
where
\begin{equation}\label{}
E_{i}=\partial_{i}A_{0} - \partial_{0}A_{i},\quad B_{3}= \partial_{1}A_{2} - \partial_{2}A_{1},\quad i=1,2,3,
\end{equation}
Then, the boundary conditions (\ref{1.27}) are satisfied if the following conditions are met:
\begin{equation}\label{1.31}
\begin{aligned}
A_{s}(x_{1},x_{2},0,\tau)=A_{s}(x_{1},x_{2},L,\tau) = 0,\quad s=0,1,2,
\end{aligned}
\end{equation}
and (\ref{1.28}) provides us with additional conditions
\begin{equation}\label{1.37}
\begin{aligned}
\partial_{3}A_{3}(x_{1},x_{2},0,\tau)=\partial_{3}A_{3}(x_{1},x_{2},L,\tau) = 0,
\end{aligned}
\end{equation}
\begin{equation}\label{1.32}
\begin{aligned}
\partial^{2}_{3}A_{0}(x_{1},x_{2},0,\tau)=\partial^{2}_{3}A_{0}(x_{1},x_{2},L,\tau) = 0.
\end{aligned}
\end{equation}
The partition function for the electromagnetic field before fixing the gauge has the form
\begin{equation}\label{1.30}
\begin{aligned}
Z = \oint \mathscr{D}A \exp\left(\int\limits_{0}^{\beta}d\tau\int\limits_{V} d^3\mathbf{x} \; \mathscr{L}_{\text{E}}\right),
\end{aligned}
\end{equation}
where the symbol $\oint$ means the functional integration taking into account periodic boundary conditions
\begin{equation}\label{1.29}
A_a(0,\mathbf{x})=A_a(\beta, \mathbf{x}),
\end{equation}
$V$ is the volume between the walls, $\beta=1/T$ is the inverse temperature, and the Euclidean action is introduced
\begin{equation}\label{}
\mathscr{L}_{\text{E}} = -\frac{1}{4} F_{ab}F_{ab},
\end{equation}
or in field strength terms
\begin{equation}\label{}
\mathscr{L}_{\text{E}} = -\frac{1}{2} \bigg(\mathbf{E}^{2}+\mathbf{B}^{2}\bigg).
\end{equation}
Let us note that in (\ref{1.30}) functional integration is not performed over the whole function space, but only over its subspace. We do not integrate over all functions defined on $V$, but only over such functions that satisfy the boundary conditions (\ref{1.31}--\ref{1.32}) as well as the periodicity conditions (\ref{1.29}). Thus, we must construct a complete orthonormal system of functions on $V$ satisfying these boundary conditions and decompose the fields by these basis functions
\begin{equation}\label{1.33}
\begin{aligned}
A_{s}(x)=\sum^{\infty}_{l=1}\sum_{n\in Z} \int d^{2}\textbf{k}_{||} \; a_{s}(\textbf{k}_{||},n,l)f(x,\textbf{k}_{||},n,l) + H.c.,\quad s=0,1,2,\\
l = 1,2\ldots ,\quad n = 0,\pm 1,\pm 2, \ldots ,
\end{aligned}
\end{equation}

\begin{equation}\label{1.34}
\begin{aligned}
A_{3}(x)=\sum^{\infty}_{l=0}\sum_{n\in Z} \int d^{2}\textbf{k}_{||} a_{3}(\textbf{k}_{||},n,l)u(x,\textbf{k}_{||},n,l) + H.c.,\\
l = 0,1,2\ldots.
\end{aligned}
\end{equation}
The boundary conditions (\ref{1.31}--\ref{1.32}) and (\ref{1.29}) are satisfied if
\begin{equation}\label{}
\begin{aligned}
f(x,\textbf{k}_{||},n,l) = \frac{\sin(q_{l}x_{3})}{\pi\sqrt{2\beta L}} e^{i(\omega_{n}\tau+\textbf{k}_{||}\textbf{x}_{||})},\\
q_{l}=\frac{\pi l}{L},\quad \omega_{n}=\frac{2\pi n}{\beta},\quad l = 0,1,2\ldots,\quad \textbf{x}_{||}=(x_1,x_2,0),\quad \textbf{k}_{||}=(k_1,k_2,0),
\end{aligned}
\end{equation}
\begin{equation}\label{}
\begin{aligned}
u(x,\textbf{k}_{||},n,l) = \frac{\cos(q_{l}x_{3})}{\pi\sqrt{2\beta L}}e^{i(\omega_{n}\tau+\textbf{k}_{||}\textbf{x}_{||})},\quad
l = 1,2\ldots,
\end{aligned}
\end{equation}
\begin{equation}\label{}
\begin{aligned}
u(x,\textbf{k}_{||},n,0) = \frac{1}{2\pi\sqrt{\beta L}}e^{i(\omega_{n}\tau+\textbf{k}_{||}\textbf{x}_{||})}.
\end{aligned}
\end{equation}
These basis functions really satisfy the completeness and normalization conditions (see Appendix \ref{Eigenfunctions}). The expression (\ref{1.30}) is invariant with respect to gauge transformations
\begin{equation}\label{1.38}
A^{\theta}_{a} = A_{a} - \partial_{a}\theta,
\end{equation}
if $\theta(x)$ satisfies the boundary conditions
\begin{equation}\label{1.39}
\begin{aligned}
\theta(\textbf{x},0)=\theta(\textbf{x},\beta),
\end{aligned}
\end{equation}
\begin{equation}\label{1.35}
\begin{aligned}
\theta(x_{1},x_{2},0,\tau)=\theta(x_{1},x_{2},L,\tau) = 0,
\end{aligned}
\end{equation}
\begin{equation}\label{1.36}
\begin{aligned}
\partial^{2}_{3}\theta(x_{1},x_{2},0,\tau)=\partial^{2}_{3}\theta(x_{1},x_{2},L,\tau) = 0.
\end{aligned}
\end{equation}
Indeed, from (\ref{1.35}) it follows that $A^{\theta}_{s}$, $s=0,1,2,$ satisfies the equality (\ref{1.31}). And it immediately follows from (\ref{1.36}) that $A^{\theta}_{3}$ satisfies (\ref{1.37}), and $A^{\theta}_{0}$ satisfies (\ref{1.32}). And it follows obviously from (\ref{1.39}) that $A^{\theta}_{a}$, $a=0,1,2,3$ satisfies (\ref{1.29}). Thus equations (\ref{1.31}--\ref{1.32}) and (\ref{1.29}) are invariant with respect to gauge transformations (\ref{1.38}) under the condition (\ref{1.39}--\ref{1.36}).
Thus in (\ref{1.30}) the functional integration is performed over the set of physically equivalent field configurations.

To eliminate unnecessary degrees of freedom, we will use the aforementioned Faddev-Popov method. Let us introduce the condition of gauge fixation
\begin{equation}\label{}
\begin{aligned}
F(A^{\theta}_{a}) = \partial_{a}A^{\theta}_{a}(x) - B(x) = 0
\end{aligned}
\end{equation}
and consider the identity
\begin{equation}\label{1.40}
\begin{aligned}
\int \mathscr{D}\theta \, \Delta_{F}(A_{a}) \, \delta [F(A^{\theta}_{a})] = 1,
\end{aligned}
\end{equation}
where
\begin{equation}\label{}
\Delta_{F}(A_{a}) = \text{Det}\bigg(\frac{\delta F(A^{\theta}_{a})}{\delta \theta}\bigg).
\end{equation}
The determinant $\Delta_{F}(A_{a})$ is gauge invariant, i.e.
\begin{equation}\label{}
\Delta_{F}(A^{\theta'}_{a}) = \Delta_{F}(A_{a}),
\end{equation}
where
\begin{equation}\label{}
\Delta_{F}(A^{\theta'}_{a}) = \text{Det}\bigg(\frac{\delta F(A^{\theta'\theta}_{a})}{\delta \theta}\bigg),
\end{equation}
\begin{equation}\label{}
A^{\theta'\theta}_{a} = A_{a} - \partial_{a}\theta' - \partial_{a}\theta.
\end{equation}
In fact, in our case $\Delta_{F}$ does not depend on $A_{a}$ at all. The functional differential $\mathscr{D}\theta$ is obviously gauge invariant as well
\begin{equation}\label{}
\mathscr{D}\theta = \mathscr{D}(\theta + \theta').
\end{equation}
Thus, the expression (\ref{1.40}) is gauge invariant. We can substitute this expression into (\ref{1.30}), since multiplication by 1 does not change the partition function
\begin{equation}\label{}
\begin{aligned}
Z = \oint \mathscr{D}A \int \mathscr{D}\theta \, \Delta_{F} \, \delta [F(A^{\theta}_{a})] \exp\left(\int\limits_{0}^{\beta}d\tau\int\limits_{V} d^3\mathbf{x} \; \mathscr{L}_{\text{E}}(A_{a})\right).
\end{aligned}
\end{equation}
Then taking into account that the Lagrangian and the field measure are also gauge invariant
\begin{equation}\label{}
\mathscr{L}_{\text{E}}(A^{\theta}_{a}) = \mathscr{L}_{\text{E}}(A_{a}),\quad
\mathscr{D}A^{\theta} = \mathscr{D}A,
\end{equation}
and changing the order of integration on $A^{\theta}_{a}$ and on $\theta$, we obtain
\begin{equation}\label{}
\begin{aligned}
Z = \int \mathscr{D}\theta \oint \mathscr{D}A^{\theta} \, \Delta_{F} \, \delta [F(A^{\theta}_{a})] \exp\left(\int\limits_{0}^{\beta}d\tau\int\limits_{V} d^3\mathbf{x} \; \mathscr{L}_{\text{E}}(A^{\theta}_{a})\right).
\end{aligned}
\end{equation}
We can always replace the dummy variable of integration $A^{\theta}_{a}$ by $A_{a}$ since the integration is carried out over all gauge-equivalent configurations
\begin{equation}\label{1.41}
\begin{aligned}
Z =\int \mathscr{D}\theta \oint \mathscr{D}A \, \Delta_{F} \, \delta [F(A_{a})] \exp\left(\int\limits_{0}^{\beta}d\tau\int\limits_{V} d^3\mathbf{x} \; \mathscr{L}_{\text{E}}(A_{a})\right).
\end{aligned}
\end{equation}
Here, functional integration over $\theta$ is equivalent to multiplying the integral by the 'volume' of the group $\int \mathscr{D}\theta$, since nothing in the integrand depends on $\theta$. Multiplication by a constant number does not affect the result, since if we calculate the mean value, this number will appear in both the numerator and the denominator, and will cancel out. Therefore, this constant can be ignored. Note that the result of computing the integral in (\ref{1.40}) does not depend on $B(x)$ at all, so we can integrate (\ref{1.41}) over $B(x)$ with additional weight $\exp (-\alpha B^2(x)/2)$
\begin{equation}\label{}
\begin{aligned}
Z =\int \mathscr{D}B\exp \bigg(-\frac{\alpha B^{2}}{2}\bigg)\oint \mathscr{D}A \, \Delta_{F} \,
\delta [\partial_{a}A_{a} - B] \; \times\\
\times \;\exp\left(\int\limits_{0}^{\beta}d\tau\int\limits_{V} d^3\mathbf{x} \; \mathscr{L}_{\text{E}}(A_{a})\right).
\end{aligned}
\end{equation}
Such integration is also equivalent to multiplication by (infinite) number. The result is
\begin{equation}\label{1.42}
\begin{aligned}
Z =\oint \mathscr{D}A \, \Delta_{F} \;\exp\left(\int\limits_{0}^{\beta}d\tau\int\limits_{V} d^3\mathbf{x} \; \mathscr{L}_{\text{eff}}\right),
\end{aligned}
\end{equation}
\begin{equation}\label{1.2}
\begin{aligned}
\mathscr{L}_{\text{eff}} = \frac{1}{2}A_{a}\square A_{a} + \frac{1-\alpha}{2}\bigg(\partial_{a}A_{a}\bigg)^{2}.
\end{aligned}
\end{equation}
Now we consider the Feynman gauge $\alpha=1$. Note that we use the Euclidean metric for D'Alambert operator, i.e. $\square = \partial_{a}\partial_{a} = \partial_{\tau}^{2} + \Delta$, $a=0,1,2,3.$
and unit system, where $k_{B}=\hbar=c=1$ ($k_B$ is the Boltzmann constant, $\hbar$ is the Planck constant, and $c$ is the speed of light).
We note that $\Delta_{F}$ can be represented in the form of
\begin{equation}\label{}
\Delta_{F}=\text{Det}\bigg(-\frac{\delta \square \theta}{\delta \theta}\bigg),
\end{equation}
where $\text{Det}(\cdot)$ is the functional determinant of operator~\cite{zinn2010path,budkov2024statistical}. A very important clarification is that here $\square$ acts on a function $\theta(x)$ satisfying quite specific fixed boundary conditions, which in turn determines the spectral properties of this operator.
In fact, it is easy to check that $\theta(x)$ satisfies the same boundary conditions as the fields $A_{s}(x)$, $s = 0,1,2,$ represented in the form (\ref{1.33}), so $\theta(x)$ can be decomposed by the same basis
\begin{equation}\label{}
\theta(x)=\sum^{\infty}_{l=1}\sum_{n\in Z} \int d^{2}\textbf{k}_{||} \; \bar{\theta}(\textbf{k}_{||},n,l)f(x,\textbf{k}_{||},n,l) + H.c.
\end{equation}
The functions $f(x,\textbf{k}_{||},n,l)$ are eigenfunctions of the operator $(-\square)$
\begin{equation}\label{}
-\square f(x,\textbf{k}_{||},n,l) = \lambda(\textbf{k}_{||},n,l) f(x,\textbf{k}_{||},n,l)
\end{equation}
with positive eigenvalues \footnote{as it should be, otherwise the integral (\ref{1.42}) would not converge, because in (\ref{1.2}) there is a plus sign before $\square$}
\begin{equation}\label{1.43}
\begin{aligned}
\lambda(\textbf{k}_{||},n,l) = \textbf{k}^{2}_{||} + q_{l}^2 +\omega_n^2,\\
l = 1,2\ldots ,\quad n = 0,\pm 1,\pm 2, \ldots 
\end{aligned}
\end{equation}
Moving from $\textbf{k}_{||}$ to a discrete spectrum and assuming that the plate sizes along $x_1$ and $x_2$, respectively, $L_x$ and $L_y$
\begin{equation}\label{}
k_1 = \frac{2\pi n_{x}}{L_x} , \quad k_2 = \frac{2\pi n_{y}}{L_y},\quad n_{x},n_{y} = 0,\pm 1,\pm 2 \ldots,
\end{equation}
and then returning back (at large $L_{x}$ and $L_{y}$) to the continuous spectrum by substitution
\begin{equation}\label{}
\sum_{n_{x}n_{y}\in Z} \rightarrow L_{x}L_{y}\int \frac{d^{2}\textbf{k}_{||}}{(2\pi)^{2}}
\end{equation}
one gets
\begin{equation}\label{1.44}
\Delta_{F}=\exp\bigg( \sum^{\infty}_{l=1} \ln\bar{\lambda}_{l}\bigg),
\end{equation}
where the following designation
\begin{equation}\label{}
\ln\bar{\lambda}_{l} = L_{x}L_{y}\int \frac{d^{2}\textbf{k}_{||}}{(2\pi)^{2}} \sum_{n\in Z} \ln \lambda(\textbf{k}_{||},n,l)
\end{equation}
is introduced. Thus, for example, $A_{0}$ and $\theta$ satisfy the same boundary conditions, and the following equality is true:
\begin{equation}\label{A0}
\begin{aligned}
\int \mathscr{D}A_{0} \, \exp\bigg(\frac{1}{2} \int\limits_{0}^{\beta}d\tau\int\limits_{V} d^3\mathbf{x} \; A_{0}\square A_{0}\bigg)=\Delta_{F}^{-\frac{1}{2}}.
\end{aligned}
\end{equation}
The same holds for $A_{1}$ and $A_{2}$. The functions $u(x,\textbf{k}_{||},n,l)$  are also eigenfunctions of the operator $(-\square)$ and for $A_3$ the following equality is satisfied
\begin{equation}\label{A3}
\begin{aligned}
\int \mathscr{D}A_{3} \, \exp\bigg(\frac{1}{2} \int\limits_{0}^{\beta}d\tau\int\limits_{V} d^3\mathbf{x} \; A_{3}\square A_{3}\bigg)=\Delta_{3}^{-\frac{1}{2}},
\end{aligned}
\end{equation}
where
\begin{equation}\label{}
\Delta_{3}=\exp\bigg( \sum^{\infty}_{l=0} \ln\bar{\lambda}_{l}\bigg).
\end{equation}
The summation over $l$ starts from $0$ instead of $1$, in contrast to (\ref{1.44}). Thus, the spectrum contains an additional set of eigenvalues corresponding to $l=0$ which are not present in the spectrum of (\ref{1.43})
\begin{equation}\label{}
\begin{aligned}
\lambda(\textbf{k}_{||},n,0) = \textbf{k}^{2}_{||} +\omega_n^{2},
\end{aligned}
\end{equation}
due to the fact that there exists a whole set of eigenfunctions that do not vanish at $l=0$
\begin{equation}\label{}
\begin{aligned}
u(x,\textbf{k}_{||},n,0) = \frac{1}{2\pi\sqrt{\beta L}}e^{i(\omega_{n}\tau+\textbf{k}_{||}\textbf{x}_{||})}.
\end{aligned}
\end{equation}
Therefore, due to different boundary conditions, the Green's functions for different fields may have different forms. In particular, due to different boundary conditions for $A_0$ and $A_3$, the operators with integral kernels $\mathscr{D}_{00}(x,y)$ and $\mathscr{D}_{33}(x,y)$ have different spectra of eigenvalues, namely, $\lambda^{-1}(\textbf{k}_{||},n,l)$, $l=1,2,3, \ldots$ for $\mathscr{D}_{00}(x,y)$ and $l=0,1,2,3, \ldots$ for $\mathscr{D}_{33}(x,y)$. The functional determinant $\Delta_{F}$ can be expressed as follows
\begin{equation}\label{Delta_F}
\Delta_{F} = \oint \mathscr{D}C^{*}\mathscr{D}C \, \exp\left(\int\limits_{0}^{\beta}d\tau\int d^3\mathbf{x} \; \partial_{a}C^{*}(x) \partial_{a}C(x)\right),
\end{equation}
where $C^{*}(x)$, $C(x)$ are the anticommuting fields that transform as scalars under the ISO(4) group transformations. Unlike commuting fields, these fields are elements of the Grassmann algebra~\cite{zinn2021quantum,weinberg1995quantum}. Functional integration, analogous to that used for the electromagnetic field, is carried out while taking into account periodic boundary conditions
\begin{equation}\label{}
C(0,\mathbf{x})=C(\beta,\mathbf{x}),\quad  C^{*}(0,\mathbf{x})=C^{*}(\beta,\mathbf{x}),
\end{equation}
therefore, these auxiliary field variables cannot be related to some real physical field, since true anticommuting fermion fields must satisfy antiperiodic boundary conditions $\psi(0,\mathbf{x})=-\psi(\beta,\mathbf{x})$~\cite{kapusta2007finite,weinberg1995quantum}. 
Thus, to satisfy the identity (\ref{Delta_F}) more exactly one can put that the ghosts satisfy the same boundary conditions as $\theta(x)$, respectively
\begin{equation}\label{1.59}
\begin{aligned}
C(x)=\int d^{2}\textbf{k}_{||} \; \sum^{\infty}_{l=1}\sum_{n\in Z} \; \eta(\textbf{k}_{||},n,l)f(x,\textbf{k}_{||},n,l),
\end{aligned}
\end{equation}
where $\eta(\textbf{k}_{||},n,l)$ are the Grassmann numbers. {Thus, from (\ref{1.44}),  (\ref{A0}) and (\ref{A3})
one gets the partition function
\begin{equation}\label{}
 Z =\exp\bigg[ - L_xL_y \int\frac{d^{2}\textbf{k}_{||}}{(2\pi)^{2}} \;
\sum_{l = 0}^{\infty}{}' \; \sum_{n \in Z} \;
\ln \bigg(\omega_{n}^{2} + \textbf{k}^{2}_{||} + q_l^2 \bigg)\bigg].
\end{equation}
}

\section{Helmholtz free energy}
In the previous section, we derived the formula for the partition function; therefore, we can now proceed to calculate the Helmholtz free energy. Note that the Helmholtz free energy is
\begin{equation}\label{Helmholtz}
F(L)=-\frac{1}{\beta }\ln {Z(L)}.
\end{equation}
Let us introduce the auxiliary function
\begin{equation}\label{free energy}
 f(\varepsilon, L) = \frac{1}{\beta} \int\frac{d^{2}\textbf{k}_{||}}{(2\pi)^{2}} \;
\sum_{l = 0}^{\infty}{}' \; \sum_{n \in Z} \;
\ln \bigg(\omega_{n}^{2} + \varphi_l^2 \bigg),
\end{equation}
where
\begin{equation}\label{}
\varphi_l = \sqrt{\textbf{k}^{2}_{||} + q_l^2 + \varepsilon}.
\end{equation}
then
\begin{equation}\label{free energy2}
 f(0, L)=\frac{F}{L_xL_y}, 
\end{equation}
is the Helmholtz free energy of the equilibrium quantum electromagnetic field per unit area of the conductive walls of a slit-like pore. 
In the following, one can use the formula. 
\begin{equation}\label{1.78}
\sum_{n \in Z}\mathscr{U}(n)=\frac{1}{2i} \oint\limits_{C} dz \; \text{ctg} (\pi z) \mathscr{U}(z),
\end{equation}
where integration is performed over the contour, $C$, containing all integer poles $z=0,\pm1,\pm2,\ldots$, $\mathscr{U}(z)$ - arbitrary function that does not have poles on the real axis. 
Using eq. (\ref{1.78}), one gets
\begin{equation}\label{1.84}
\sum_{n \in Z} \;\frac{1}{\omega_{n}^{2} + \varphi_l^2} =  \frac{\beta}{2\varphi_l}\;\text{cth}\bigg(\frac{\beta \varphi_l}{2}\bigg),
\end{equation}
thus
\begin{equation}\label{}
\frac{\partial f}{\partial \varepsilon} = \frac{1}{2} \int\frac{d^{2}\textbf{k}_{||}}{(2\pi)^{2}} \;
\sum_{l = 0}^{\infty}{'} \frac{1}{\varphi_l}\;\text{cth}\bigg(\frac{\beta \varphi_l}{2}\bigg)
\end{equation}
that in turn can be rewritten as follows
\begin{equation}\label{1.85}
\frac{\partial f}{\partial \varepsilon} =  \int\frac{d^{2}\textbf{k}_{||}}{(2\pi)^{2}} \;
\sum_{l = 0}^{\infty}{}' \; \frac{1}{2\varphi_l}\bigg(1 + \frac{2}{e^{\beta \varphi_l} - 1}
\bigg).
\end{equation}
From this, calculating the indefinite integral over $\varepsilon$, we can derive the following
\begin{equation}\label{}
f(0,L) =  f_0(L) + f_1(L),
\end{equation}
where
\begin{equation}\label{f_0}
f_0(L) =  \int\frac{d^{2}\textbf{k}_{||}}{(2\pi)^{2}} \;
\sum_{l = 0}^{\infty}{}' \; \sqrt{k^{2}_{||} + q_l^2},
\end{equation}
\begin{equation}\label{1.80}
f_1(L) =  \frac{2}{\beta} \int\frac{d^{2}\textbf{k}_{||}}{(2\pi)^{2}} \;
\sum_{l = 0}^{\infty}{'} \; 
\ln\bigg(1 - \exp\bigg(-\beta \sqrt{k^{2}_{||} + q_l^2}\bigg) \bigg).
\end{equation}
Let us introduce the function
\begin{equation}\label{cutoff sum}
U(l) = \int\frac{d^{2}\textbf{k}_{||}}{(2\pi)^{2}} \; \sqrt{k^{2}_{||} + q_l^2} \; \Phi(\rho_l^2),
\end{equation}
with
\begin{equation}\label{}
\rho_l^2 = (k^{2}_{||} +  q_l^2)/\Lambda^2,
\end{equation}
where the cutoff function $\Phi$ {satisfies identities}   
\begin{equation}\label{1.79}
\begin{aligned}
\Phi(0)=1,
\quad \lim_{\xi\rightarrow \infty}\Phi(\xi)=0.
\end{aligned}
\end{equation}
If at $z=\infty$ the function $\Phi(z^2)$ approaches zero faster than $O(1/|z|)$. Thus one can put
\begin{equation}\label{}
f_0(L) = \sum_{l = 0}^{\infty}{'} U(l).
\end{equation}
Then, using the Euler-Maclaurin formula~\cite{andrews1999special},
\begin{equation}\label{EulerMac1}
\sum_{l = 0}^{\infty} \; U(l) - \int\limits_{0}^{\infty} d\xi \; U(\xi) =
\frac{U(0)}{2} - i\int\limits_{0}^{\infty} d\xi \;\frac{U(i\xi)-U(-i\xi)}{e^{2\pi\xi}-1},
\end{equation}
we obtain
\begin{equation}\label{1.83}
\lim_{\Lambda\rightarrow \infty}\sum_{l = 0}^{\infty}{'} \; U(l)= - \frac{\pi^2}{720 L^3}
+\lim_{\Lambda\rightarrow \infty}\int\limits_{0}^{\infty} d\xi \; U(\xi).
\end{equation} 
The second term on the right-hand side of (\ref{1.83}) can be more precisely defined as
\begin{equation}\label{1.88}
\begin{aligned}
\lim_{\Lambda\rightarrow \infty}\int\limits_{0}^{\infty} d\xi \; U(\xi)  = 
\frac{\mathscr{E}_{vac}}{L_xL_y},
\end{aligned}
\end{equation}\label{}
where the vacuum energy density is introduced
\begin{equation}
\mathscr{E}_{vac} = V\lim_{a\rightarrow \infty} \int\limits_{0}^{a} \frac{dk}{2\pi^{2}} \bigg(\;k^3 \; \Phi\bigg(k^2/\Lambda^2(a)\bigg)\bigg),\quad V=LL_xL_y,
\end{equation}
\begin{equation}\label{}
\lim_{a\rightarrow \infty}\Lambda(a)= \infty,  
\end{equation}
that in SI units is
\begin{equation}\label{}
\begin{aligned}
\mathscr{E}_{vac} = sV\lim_{a\rightarrow \infty} \int \frac{d^3\textbf{k}}{(2\pi)^{3}} \; 
\bigg(\frac{\hbar \bar{\omega}}{2}\; \Phi\bigg(k^2/\Lambda^2(a)\bigg)\bigg),
\end{aligned}
\end{equation}
\begin{equation}\label{}
\bar{\omega} = k c,    
\end{equation}
\begin{equation}\label{}
\begin{aligned}
\int \frac{d^3\textbf{k}}{(2\pi)^{3}} \; 
\bigg(\hbar \bar{\omega}\; \Phi\bigg(k^2/\Lambda^2(a)\bigg)\bigg) =\hbar c \int \frac{dk}{2\pi^{2}} \bigg(k^3\; \Phi\bigg(k^2/\Lambda^2(a)\bigg)\bigg),
\end{aligned}
\end{equation}
where $s=2$ is the number of possible polarizations of the photon~\cite{berestetskii1982quantum}.\par
{Then substituting (\ref{f_0}) and (\ref{1.80}) into equation (\ref{Helmholtz}) with account of (\ref{1.83}) one can finally write the following 
\begin{equation}\label{1.98}
\begin{aligned}
F =  F_0  + F_1, 
\end{aligned}
\end{equation}
\begin{equation}\label{}
F_0 =  L_xL_y f_0,\quad F_1 =  L_xL_y f_1,       
\end{equation}
where
\begin{equation}\label{1.99}
F_0 = \mathscr{E}_{vac}  - \frac{\pi^2 L_xL_y}{720 L^3},  
\end{equation}
\begin{equation}\label{2.00}
F_1 = \frac{2 L_xL_y}{\beta} \int\frac{d^{2}\textbf{k}_{||}}{(2\pi)^{2}} \;
\sum_{l = 0}^{\infty}{'} \; 
\ln\bigg(1 - \exp\bigg(-\beta \sqrt{k^{2}_{||} + q_l^2}\bigg) \bigg).    
\end{equation}
}

\section{Casimir stresses}\label{CF}

\subsection{Casimir stress tensor}
\subsubsection{General case}
Taking into account (\ref{Delta_F}), the partition function (\ref{1.42}) takes the form
\begin{equation}\label{}
\begin{aligned}
Z = \oint \mathscr{D}A\mathscr{D}C^{*}\mathscr{D}C \; \exp\left(\int\limits_{0}^{\beta}d\tau\int d^3\mathbf{x} \; \mathscr{L}\right),
\end{aligned}
\end{equation}
\begin{equation}\label{1.51}
\mathscr{L} = - \frac{1}{2} \partial_{b}A_{a}(x) \partial_{b}A_{a}(x) + \partial_{a}C^{*}(x) \partial_{a}C(x).
\end{equation}
Consequently, the Noether current tensor is 
\begin{equation}\label{1.48}
\begin{aligned}
T_{ab} = - \partial_{a}A_{c}\partial_{b}A_{c} + \frac{1}{2} \delta_{ab}\partial_{d}A_{c}\partial_{d}A_{c}+ \delta_{ab} \partial_d C\partial_d C^{*}\\
- \partial_a C\partial_b C^{*} - \partial_b C\partial_a C^{*}.
\end{aligned}
\end{equation}
As it follows from the thermomechanical approach based on the Noether's first theorem~\cite{budkov2024thermomechanical,brandyshev2023statistical}, the conservation low is
\begin{equation}\label{}
\partial_{a} \bigg\langle T_{ab}(x)\bigg\rangle = 0,
\end{equation}
where the average energy-momentum tensor is determined up to divergenceless terms as follows
\begin{equation}\label{1.61}
\begin{aligned}
\bigg\langle T_{ab} \bigg\rangle = \lim_{x'\rightarrow x}\bigg[
\bigg(\delta_{ab}\partial_{d}'\partial_{d} - \partial_{a}'\partial_{b} - \partial_{a}\partial_{b}'\bigg)
\bigg(\frac{1}{2} \mathscr{D}_{cc}(x,x') + \mathscr{D}_{C}(x,x')\bigg)\bigg],
\end{aligned}
\end{equation}
where the average is defined by 
\begin{equation}
 \bigg\langle\bigg(\cdot\bigg)\bigg\rangle=Z^{-1}\oint \mathscr{D}A\,\mathscr{D}C^{*}\mathscr{D}C \,\exp\left(\int\limits_{0}^{\beta}d\tau\int d^3\mathbf{x} \; \mathscr{L}\right) \bigg(\cdot\bigg).     
\end{equation}
The Green's function for the electromagnetic field, $\mathscr{D}_{ab}(x,y)=\bigg\langle A_a(x)A_b(y)\bigg\rangle$, is defined by the standard equation
\begin{equation}\label{1.67}
-\square\mathscr{D}_{ab}(x,y)=\delta_{ab}\delta(x-y),
\end{equation}
and the Green's function for the ghost field, $\mathscr{D}_{C}(x,y)= \bigg\langle C(x)C^{*}(y)\bigg\rangle$, is defined by
\begin{equation}\label{1.25}
\square\mathscr{D}_{C}(x,y)=\delta(x-y).
\end{equation}
The Casimir stress tensor represents the spatial components of the energy-momentum tensor, $T_{ij}$, $i,j=1,2,3$. Consequently, the average value of the mechanical stress tensor is
\begin{equation}\label{stress}
\sigma_{ij} = \bigg\langle T_{ij} \bigg\rangle.
\end{equation}

We would like to draw the reader's attention that the approaches based on applying the stress tensor to describe the Casimir effect have been developed by other authors \cite{dzyaloshinskii1961general,schwinger1978casimir,bordag1985quantum,brown1969vacuum,tadaki1986casimir}. However, our work introduces a novel aspect by applying Noether's theorem~\cite{brandyshev2023noether,brandyshev2023statistical,hermann2022noether} to the partition function of the electromagnetic field in a functional integral form, which is new compared to other approaches. Note also that the general stress tensor (\ref{stress}) can be used to calculate the Casimir force between metals of arbitrary shape in the approximation of an ideal conductor by the expression $F_{i}=\oint\limits dS n_{j}\sigma_{ij}$, where the integration is carried out over the surface of the body, with $\mathbf{n}$ representing the normal vector. In this paper, we will not consider the case of bodies with a curved surface, but rather limit ourselves to verifying the correctness of the stress tensor by applying it to the case of a slit-like pore. For this case, an expression for the normal stress (Casimir force) has been obtained by other methods, both at zero temperature and considering thermal corrections for radiation. Additionally, we will calculate a previously unknown tangential stress.

\subsubsection{Slit-like pore case}
According to the boundary conditions (\ref{1.31}--\ref{1.32})
\begin{equation}\label{}
\mathscr{D}_{00}(x,y)=\mathscr{D}_{11}(x,y)=\mathscr{D}_{22}(x,y)=\mathscr{D}(x,y),
\end{equation}
where the operator $\mathscr{D}(x,y)$ is determined by the equation
\begin{equation}\label{1.60}
-\square\mathscr{D}(x,y)=\delta(x-y)
\end{equation}
with account of the boundary conditions
\begin{equation}\label{}
\mathscr{D}(\tau,x_{1},x_{2},0;\;y)=\mathscr{D}(\tau,x_{1},x_{2},L;\;y)=0,
\end{equation}
\begin{equation}\label{}
\mathscr{D}(0,\textbf{x};\;y)=\mathscr{D}(\beta,\textbf{x};\;y).
\end{equation}
For the operator $\mathscr{D}_{33}$, an equation analogous to (\ref{1.60}) holds
\begin{equation}\label{}
-\square\mathscr{D}_{33}(x,y)=\delta(x-y),
\end{equation}
but with different boundary conditions
\begin{equation}\label{}
\partial_{3}\mathscr{D}_{33}(\tau,x_{1},x_{2},0;\;y)=\partial_{3}\mathscr{D}_{33}(\tau,x_{1},x_{2},L;\;y)=0,
\end{equation}
\begin{equation}\label{}
\mathscr{D}_{33}(0,\textbf{x};\;y)=\mathscr{D}_{33}(\beta,\textbf{x};\;y).
\end{equation}
The Green's function for the ghost fields $\mathscr{D}_{C}$ satisfies the equation (\ref{1.25}). From (\ref{1.59}) it follows that $\mathscr{D}_{C}$ has the same boundary conditions as the $\mathscr{D}$; therefore, by comparing (\ref{1.60}) with (\ref{1.25}), it can be seen
\begin{equation}\label{}
\mathscr{D}_{C}(x,y) = - \mathscr{D}(x,y),
\end{equation}
and of course, from (\ref{1.67}), it follows that
\begin{equation}\label{}
\mathscr{D}_{ab}(x,y)=0,\quad a\neq b.
\end{equation}
Solutions of these equations have the form
\begin{equation}\label{1.65}
\begin{aligned}
\mathscr{D}(x,y)=\frac{2}{\beta L}\int\frac{d^{2}\textbf{k}_{||}}{(2\pi)^{2}} \;
\sum_{n \in Z}\sum^{\infty}_{l=1} \;
\frac{\sin(q_{l}x_{3})\sin(q_{l}x'_{3})}{\textbf{k}^{2}_{||} + q^{2}_{l} + \omega_{n}^{2}}e^{i\omega_{n}(\tau-\tau')+i\textbf{k}_{||}(\textbf{x}_{||}-\textbf{x}'_{||})},
\end{aligned}
\end{equation}
\begin{equation}\label{1.66}
\begin{aligned}
\mathscr{D}_{33}(x,y)=\frac{2}{\beta L}\int\frac{d^{2}\textbf{k}_{||}}{(2\pi)^{2}} \;
\sum_{n \in Z}\sum^{\infty}_{l=0}{}' \;
\frac{\cos(q_{l}x_{3})\cos(q_{l}x'_{3})}{\textbf{k}^{2}_{||} + q^{2}_{l} + \omega_{n}^{2}}
e^{i\omega_{n}(\tau-\tau')+i\textbf{k}_{||}(\textbf{x}_{||}-\textbf{x}'_{||})}.
\end{aligned}
\end{equation}
A dashed sum means that the term with the number $l=0$ is included with a multiplier $1/2$. The expression (\ref{1.65}) is obtained by substituting the formulas (\ref{1.62}) and (\ref{1.64}) from the Appendix \ref{Eigenfunctions} into (\ref{1.60}), similarly (\ref{1.66}) is obtained by substituting the formulas (\ref{1.63}) and (\ref{1.64}) (see also Appendix \ref{Eigenfunctions}) into (\ref{1.60}).\par 

Thus from (\ref{1.61}) one gets
\begin{equation}\label{1.73}
\begin{aligned}
\bigg\langle T_{ab}\bigg\rangle = \frac{1}{2}\lim_{x'\rightarrow x}\bigg[
\bigg(\delta_{ab}\partial_{c}'\partial_{c} - \partial_{a}'\partial_{b} - \partial_{a}\partial_{b}'\bigg)
\bigg(\mathscr{D}(x,x') + \mathscr{D}_{33}(x,x')\bigg)\bigg].
\end{aligned}
\end{equation}
\subsection{Expansion near the classical limit ($\omega_n=0$)}
\subsubsection{Normal stress and disjoining pressure}
Let us consider the normal component of the stress tensor relative to the surface of the walls that defines the normal pressure
\begin{equation}\label{}
\sigma_{33} = \bigg\langle T_{33} \bigg\rangle,
\end{equation}
\begin{equation}\label{}
\begin{aligned}
\sigma_{33} = \lim_{x'\rightarrow x}\bigg[
\bigg(\frac{1}{2}\partial_{c}'\partial_{c} - \partial_{3}'\partial_{3}\bigg)
\bigg(\mathscr{D}(x,x') + \mathscr{D}_{33}(x,x')\bigg)\bigg],
\end{aligned}
\end{equation}
that can be rewritten as
\begin{equation}\label{}
\begin{aligned}
\sigma_{33} = \lim_{x'\rightarrow x}\bigg[
\bigg(\partial_{s}'\partial_{s} - \frac{1}{2}\partial_{c}'\partial_{c}\bigg)
\bigg(\mathscr{D}(x,x') + \mathscr{D}_{33}(x,x')\bigg)\bigg].
\end{aligned}
\end{equation}
Using (\ref{1.65}--\ref{1.66})  it is easy to show that
\begin{equation}\label{dsds}
\begin{aligned}
\partial_{s}\partial'_{s}\mathscr{D}_{ab}(x,x') = -\partial_{s}\partial_{s}\mathscr{D}_{ab}(x,x'),\\
s=0,1,2,\quad a,b=0,1,2,3,\\
\end{aligned}
\end{equation}
{\begin{equation}\label{}
\begin{aligned}
\lim_{x'\rightarrow x}\partial_{c}\partial'_{c}\bigg(\mathscr{D}(x,x') + \mathscr{D}_{33}(x,x')\bigg)=
\lim_{x'\rightarrow x}\bigg[-\square_{x}\bigg(\mathscr{D}(x,x') + \mathscr{D}_{33}(x,x')\bigg)\bigg],
\end{aligned}
\end{equation}
then
\begin{equation}\label{}
\begin{aligned}
\sigma_{33} = \lim_{x'\rightarrow x}\bigg[\bigg(\frac{1}{2}\square_{x} - \partial_{s}\partial_{s}\bigg)\bigg(\mathscr{D}(x,x')+\mathscr{D}_{33}(x,x')\bigg)\bigg].
\end{aligned}
\end{equation}
From (\ref{1.60}) it follows that
\begin{equation}\label{Delta_zero}
\partial_{a}\lim_{x'\rightarrow x}\bigg[\square_{x}\bigg(\mathscr{D}(x,x')+\mathscr{D}_{33}(x,x')\bigg)\bigg]=0.
\end{equation}
Therefore, since the normal stress, $\sigma_{33}$, is determined up to divergenceless terms, we can assume 
\begin{equation}\label{}
\sigma_{33} =
\frac{1}{\beta L} \int\frac{d^{2}\textbf{k}_{||}}{(2\pi)^{2}} \;
\sum_{n,l \in Z} \;
\frac{\textbf{k}^{2}_{||} + \omega_{n}^{2}}{\textbf{k}^{2}_{||} + q^{2}_{l} + \omega_{n}^{2}}.
\end{equation}
The normal pressure is
\begin{equation}\label{}
\text{P}_{\bot}=-\sigma_{33}=
-\frac{1}{\beta L} \int\frac{d^{2}\textbf{k}_{||}}{(2\pi)^{2}} \;
\sum_{n,l \in Z} \;
\frac{\textbf{k}^{2}_{||} + \omega_{n}^{2}}{\textbf{k}^{2}_{||} + q^{2}_{l} + \omega_{n}^{2}}.
\end{equation}
It can be written that
\begin{equation}\label{1.68}
\text{P}_{\bot}=
-\frac{1}{\beta L} \int\frac{d^{2}\textbf{k}_{||}}{(2\pi)^{2}}  \;
\sum_{n \in Z}\bigg[\bigg(\textbf{k}^{2}_{||} + \omega_{n}^{2}\bigg)\sum_{l \in Z} \; \mathscr{U}(l)\bigg],
\end{equation}
where
\begin{equation}\label{}
\mathscr{U}(l)=\frac{1}{\textbf{k}^{2}_{||} + \omega_{n}^{2} + (\pi l / L )^{2}}.
\end{equation}
Thus one can use the formula (\ref{1.78}) and then one gets
\begin{equation}\label{}
\begin{aligned}
\sum_{l \in Z}\mathscr{U}(l) = \frac{1}{2i} \oint\limits_{C} dz \; \frac{\text{ctg} (\pi z)}{(\pi / L )^{2}(z-z_{+})(z-z_{-})},\\
z_{\pm}=\pm \frac{iL}{\pi}\sqrt{\textbf{k}^{2}_{||} + \omega_{n}^{2}}.
\end{aligned}
\end{equation}
From Cauchy's theorem, it follows that the sum of all residues over the entire complex plane, including the point at infinity, is equal to zero. Therefore, considering that the poles $z_+$ and $z_-$ are located outside the contour of integration, we can change the direction of integration and obtain the following equality for arbitrary finite $L$, $k_{||}$ and $\omega_{n}$
\begin{equation}\label{sum over l}
\begin{aligned}
\sum_{l \in Z}\mathscr{U}(l) = 
\; \frac{L \,\text{cth} \bigg(L \sqrt{k^{2}_{||} + \omega_{n}^{2}}\bigg)}{\sqrt{k^{2}_{||} + \omega_{n}^{2}}}.
\end{aligned}
\end{equation}
Then substituting this sum into (\ref{1.68}), we can get}
\begin{equation}\label{1.89}
\text{P}_{\bot}(L) =
- \frac{T}{2\pi} \int\limits_{0}^{\infty} dk_{||} \; k_{||} \;
\sum_{n \in Z} \;
\sqrt{k^{2}_{||} + \omega_{n}^{2}}\; \text{cth} \bigg(L \sqrt{k^{2}_{||} + \omega_{n}^{2}}\bigg).
\end{equation}

To obtain the observed disjoining pressure (the excess pressure exerted on the wall of a pore compared to the pressure that would be exerted in the absence of walls, i.e., in a bulk system~\cite{derjaguin1987surface}), the pressure in the bulk must be subtracted from here 
\begin{equation}\label{}
\Pi(L)=\text{P}_{\bot}(L)-\text{P}_{\bot}^{(b)},\quad \text{P}_{\bot}^{(b)}=\lim_{L\rightarrow\infty}\text{P}_{\bot}(L).
\end{equation}
Therefore, we obtain the expression obtained for the first time by Schwinger et al.\cite{schwinger1978casimir}
\begin{equation}\label{1.69}
\Pi(L) =
- \frac{T}{2\pi} \int\limits_{0}^{\infty} dk_{||} \; k_{||} \;
\sum_{n \in Z} \;
\sqrt{k^{2}_{||} + \omega_{n}^{2}}\bigg(\text{cth} \bigg(L \sqrt{k^{2}_{||} + \omega_{n}^{2}}\bigg)-1\bigg).
\end{equation}

\subsubsection{Normal stress in the bulk}
Substituting the cutoff function $\Phi$ defined in (\ref{1.79}) into (\ref{1.89}), one gets
\begin{equation}\label{1.91}
\begin{aligned}
\text{P}^{(b)}_{\bot} =
- 2T \sum_{n = 0}^{\infty}{}' \; v(n),
\end{aligned}
\end{equation}
\begin{equation}\label{v(n)}
\begin{aligned}
v(n) =
\int \frac{d^2\textbf{k}_{||}}{(2\pi)^2} \;
\bigg(\sqrt{k^{2}_{||} + \omega_{n}^{2}} \;\Phi(R_n^2)\bigg),
\end{aligned}
\end{equation}
\begin{equation}\label{}
\begin{aligned}
R_n^2 = (k^{2}_{||} + \omega_{n}^{2})/\Lambda^2.
\end{aligned}
\end{equation}
Then using the Euler--Maclaurin formula by analogy with the expression (\ref{1.83}) one has
\begin{equation}\label{1.92}
\lim_{\Lambda\rightarrow \infty}\sum_{n = 0}^{\infty}{}' \; v(n)= - \frac{\pi^2 T^3}{90}
+\lim_{\Lambda\rightarrow \infty}\int\limits_{0}^{\infty} d\xi \; v(\xi),
\end{equation}
where
\begin{equation}
\lim_{\Lambda\rightarrow \infty}\int\limits_{0}^{\infty} d\xi \; v(\xi) = \frac{1}{2T}\frac{d\mathscr{E}_{vac}}{dV}.    
\end{equation}
Thus, substituting (\ref{1.92}) into (\ref{1.91}) one derives 
\begin{equation}\label{1.81}
\begin{aligned}
\text{P}_{\bot}^{(b)}
= \text{P}^{(b)}_{B} - \frac{d\mathscr{E}_{vac}}{dV},
\end{aligned}
\end{equation}
where we have introduced the pressure of the black body radiation (photon gas) in the thermodynamic limit~\cite{landau2013classical}, i.e. in the limit of $L\rightarrow\infty$, 
\begin{equation}\label{}
\begin{aligned}
\text{P}^{(b)}_{B} = \frac{\pi^2 T^4}{45},
\end{aligned}
\end{equation}
that in SI units can be written as
\begin{equation}\label{}
\begin{aligned}
\text{P}^{(b)}_{B} = \frac{\pi^2 k_B^4T^4}{45\hbar^3 c^3}.
\end{aligned}
\end{equation}
\subsubsection{Disjoining pressure at zero temperature}
The expression (\ref{1.69}) can be rewritten as 
\begin{equation}\label{2.04}
\begin{aligned}
\Pi(L) =
- \frac{T}{\pi} \sum_{n \in Z}\int\limits_{0}^{\infty}\;dk_{||} \; k_{||} \;
\frac{\sqrt{k^{2}_{||} + \omega_{n}^{2}}}{\exp\bigg(2L\sqrt{k^{2}_{||} + \omega_{n}^{2}}\bigg)-1}.
\end{aligned}
\end{equation}
In the regime of extremely small temperatures $T\rightarrow0$ (in the absence of real photons), using the substitution $2\pi T \rightarrow d\omega$ and taking into account that the integrand is an even function of $\omega$, one can show that the sum over $n$ turns into an integral and then one derives
\begin{equation}\label{}
\begin{aligned}
\Pi_{C}(L) =
- \frac{1}{\pi^{2}} \int\limits_{0}^{\infty} d\omega\int\limits_{0}^{\infty}\; dk_{||}\; k_{||} \;
\frac{\xi}{e^{2\xi L}-1},
\end{aligned}
\end{equation}
\begin{equation}\label{}
\xi = \sqrt{k^{2}_{||} + \omega^{2}}.
\end{equation}
After the transition to the polar coordinates
\begin{equation}\label{1.75}
\omega=\xi \sin \varphi,\quad k_{||} = \xi \cos \varphi,\quad d\omega dk_{||}= \xi d\xi d\varphi,
\end{equation}
we can get
\begin{equation}\label{}
\begin{aligned}
\Pi_{C}(L) =
- \frac{1}{\pi^{2}} \int\limits_{0}^{\pi/2}d\varphi \; \cos \varphi \int\limits_{0}^{\infty}\; d\xi \;
\frac{\xi^{3}}{e^{2\xi L}-1},
\end{aligned}
\end{equation}
whence with account of the well-known integral
\begin{equation}\label{1.76}
\int\limits_{0}^{\infty}\; d\xi \;
\frac{\xi^{3}}{e^{\xi}-1} = \frac{\pi^{4}}{15},
\end{equation}
finally one obtains the well-known Casimir expression~\cite{casimir1948attraction}
\begin{equation}\label{casimir_force}
\begin{aligned}
\Pi_{C}(L) = - \frac{\pi^{2}}{240 L^{4}}.
\end{aligned}
\end{equation}
\subsubsection{Disjoining pressure in the classical limit and quantum corrections}
Let us consider the zero Matsubara mode, $n=0$, and then from the expression (\ref{1.69}) we get the Casimir pressure in the classical regime~\cite{schwinger1978casimir,budkov2025first}
\begin{equation}\label{classical limit}
\begin{aligned}
\Pi_{cl}(L) =
- T \int\frac{d^{2}\textbf{k}_{||}}{(2\pi)^{2}} \;
k_{||}\bigg(\text{cth} (k_{||}L)-1\bigg)=-\frac{T\zeta(3)}{4\pi L^3}.
\end{aligned}
\end{equation}
Note that this result is for an ideal conductor with a reflection coefficient of transverse electric mode of 1. If the reflection coefficient of transverse electric mode is 0, the force would be two times smaller in the limit of a perfect conductor~\cite{buenzli2008microscopic}. This force can also be obtained for conductive walls separated by a rather large distance, modeled as the interaction between half-spaces containing an equilibrium plasma composed of classical ions~\cite{ninham1997ion,buenzli2005microscopic,budkov2023variational}. We also address these issues in our recent paper~\cite{budkov2025first}.\par
{Now we can calculate corrections to the classical limit (\ref{classical limit}). Taking the integral in the expression (\ref{1.69}) by parts one gets
\begin{equation}\label{2.06} 
\begin{aligned}
\Pi(L) = \Pi_{cl}(L)
- \frac{T}{2\pi L^3} \sum_{n = 1}^{\infty} \; \bigg( \text{Li}_{3}(e^{-2 \omega_n L}) 
+ 2 \omega_n L\;\text{Li}_{2}(e^{-2 \omega_n L})
+ 2 \omega_n^2 L^2\;\text{Li}_{1}(e^{-2 \omega_n L})\bigg),
\end{aligned}
\end{equation}
where $\text{Li}_s(x) = \sum_{k = 1}^{\infty} x^k / k^s$ is the logarithmic integral of order $s$. 
}
\subsubsection{Tangential pressure} 

Now let us obtain the tangential pressure $P_{||}=-\sigma_{11}=-\sigma_{22}$. From eq. (\ref{1.73}), it follows that $\sigma_{11}$ is determined up to divergenceless terms as follows
\begin{equation}\label{}
\sigma_{11} = - \lim_{x'\rightarrow x}\bigg[
\partial_{1}'\partial_{1}
\bigg(\mathscr{D}(x,x') + \mathscr{D}_{33}(x,x')\bigg)\bigg],
\end{equation}
thus the tangential pressure is
\begin{equation}\label{1.82}
\text{P}_{||}  = \frac{1}{\beta L}\int\frac{d^{2}\textbf{k}_{||}}{(2\pi)^{2}} \;
\sum_{n \in Z}\sum^{\infty}_{l=0}{}' \;
\frac{k^{2}_{||}}{k^{2}_{||} + q^{2}_{l} + \omega_{n}^{2}},
\end{equation}
where $k_{||}^2=k_1^2+k_2^2$ and the dashed sum means that the term corresponding to $l=0$ is included with the multiplier $1/2$ and the following equality is taken into account
\begin{equation}\label{identity1}
\int\frac{d^{2}\textbf{k}_{||}}{(2\pi)^{2}}\; \frac{k^{2}_{1}}{k^{2}_{||} + q^{2}_{l} + \omega_{n}^{2}}
=\frac{1}{2}\int\frac{d^{2}\textbf{k}_{||}}{(2\pi)^{2}}\; \frac{k^{2}_{||}}{k^{2}_{||} + q^{2}_{l} + \omega_{n}^{2}}.
\end{equation} 
Using the equation (\ref{1.78}), from (\ref{1.82}) we can obtain
\begin{equation}\label{1.90}
\text{P}_{||} = \frac{T}{2}\sum_{n \in Z}\int\frac{d^{2}\textbf{k}_{||}}{(2\pi)^{2}} \;
\frac{k^{2}_{||}\; \text{cth}\bigg(L\sqrt{k^{2}_{||} + \omega_{n}^{2}}\bigg)}{\sqrt{k^{2}_{||} + \omega_{n}^{2}}}.
\end{equation}
Then subtracting from this the pressure in the bulk 
\begin{equation}
\Pi_{||} = \text{P}_{||} - \text{P}^{(b)}_{||}, \quad  \text{P}^{(b)}_{||} = \lim_{L\rightarrow\infty}\text{P}_{||}.
\end{equation}
one can obtain the following
\begin{equation}\label{tangential stress}
\Pi_{||} = \frac{T}{2}\sum_{n \in Z}\int\frac{d^{2}\textbf{k}_{||}}{(2\pi)^{2}} \;
\frac{k^{2}_{||}\; \bigg(\text{cth}\bigg(L\sqrt{k^{2}_{||} + \omega_{n}^{2}}\bigg) - 1\bigg)}{\sqrt{k^{2}_{||} + \omega_{n}^{2}}},
\end{equation}
It can be shown that the tangential pressure in the bulk $\text{P}^{(b)}_{||}=\text{P}^{(b)}_{\bot}$ (see Appendix \ref{Tangential pressure in the bulk}).
{Thus quantum corrections to the classical tangential pressure is defined by the expression 
\begin{equation}\label{2.05}
\begin{aligned}
\Pi_{||}(L) = \frac{T\zeta(3)}{8\pi L^3}
+ \frac{T}{4\pi L^3} \sum_{n = 1}^{\infty} \; \bigg( \text{Li}_{3}(e^{-2 \omega_n L}) 
+ 2 \omega_n L\;\text{Li}_{2}(e^{-2 \omega_n L})
\bigg).
\end{aligned}
\end{equation}
This is a novel outcome of this research, akin to equation (\ref{2.06}) for the disjoining pressure, derived from the Schwinger's formula (\ref{1.69}).}

\subsubsection{Tangential pressure at zero temperature} 
The expression (\ref{tangential stress}) can be rewritten as
\begin{equation}\label{}
\Pi_{||} = T\sum_{n \in Z}\int\frac{d^{2}\textbf{k}_{||}}{(2\pi)^{2}} \;
\frac{k^{2}_{||}}{\xi_{n} \bigg(e^{2\xi_{n} L}-1\bigg)},
\end{equation}
where
\begin{equation}\label{}
\xi_{n}=\sqrt{k^{2}_{||} + \omega_{n}^{2}}.
\end{equation}
In the vacuum, i.e. at $T\rightarrow 0$ the sum turns out into the integral
\begin{equation}\label{}
\begin{aligned}
\Pi^{(C)}_{||} = 2\int\limits_{0}^{\infty} d\omega\int\limits_{0}^{\infty}\frac{dk_{||}}{(2\pi)^{2}} \;
\frac{k^{3}_{||}}{\xi \bigg(e^{2\xi L}-1\bigg)},
\end{aligned}
\end{equation}
where
\begin{equation}\label{}
\xi=\sqrt{k^{2}_{||} + \omega^{2}},
\end{equation}
and using the polar coordinates $\omega=\xi \sin\varphi$, $k_{||} = \xi \cos \varphi$, (\ref{1.75}) one derives
\begin{equation}\label{}
\begin{aligned}
\Pi^{(C)}_{||} =  2\int\limits_{0}^{\pi/2} d\varphi \; \cos^{3} \varphi \int\limits_{0}^{\infty}\frac{d \xi}{(2\pi)^{2}} \;
\frac{\xi^{3}}{e^{2\xi L}-1}.
\end{aligned}
\end{equation}
Thus, taking the integral with using the integral (\ref{1.76}), we can obtain 
\begin{equation}\label{2.03}
\begin{aligned}
\Pi^{(C)}_{||} = \frac{\pi^{2}}{720 L^{4}}.
\end{aligned}
\end{equation}

\subsection{Expansion near the vacuum limit ($T=0$)}
In this section, we will formulate and discuss a method for obtaining thermal corrections to the Casimir attractive force between conductive walls, caused by black body radiation. We will also calculate the total tangential pressure that occurs in a slit-like pore with conductive walls. 
\subsubsection{Normal pressure} 
Using (\ref{1.65}--\ref{1.66}) it can be shown that
\begin{equation}\label{}
\begin{aligned}
\lim_{x'\rightarrow x}\partial_{3}\partial'_{3}\mathscr{D}_{ab}(x,x') =
\lim_{x'\rightarrow x}\partial_{3}^{2}\mathscr{D}_{ab}(x,x')  \; + \\
+ \; \frac{2}{\beta L}\int\frac{d^{2}\textbf{k}_{||}}{(2\pi)^{2}} \;
\sum_{n \in Z}\sum^{\infty}_{l=1} \;
\frac{q^{2}_{l}}{\textbf{k}^{2}_{||} + q^{2}_{l} + \omega_{n}^{2}},
\end{aligned}
\end{equation}
then from this identity and using the equation (\ref{dsds}) one gets
\begin{equation}\label{}
\begin{aligned}
\sigma_{33} = \frac{1}{2}\lim_{x'\rightarrow x}\bigg[-\square_{x}\bigg(\mathscr{D}(x,x')+\mathscr{D}_{33}(x,x')\bigg)\bigg] - \\
- \frac{2}{\beta L}\int\frac{d^{2}\textbf{k}_{||}}{(2\pi)^{2}} \;
\sum_{n \in Z}\sum^{\infty}_{l=1} \;
\frac{q^{2}_{l}}{\textbf{k}^{2}_{||} + q^{2}_{l} + \omega_{n}^{2}}.
\end{aligned}
\end{equation}
Therefore, since the normal stress, $\sigma_{33}$, is determined up to divergenceless terms, with account of (\ref{Delta_zero}) we can assume 
\begin{equation}\label{1.77}
\sigma_{33} =
- \frac{1}{\beta L} \int\frac{d^{2}\textbf{k}_{||}}{(2\pi)^{2}} \;
\sum_{n,l \in Z} \;
\frac{q^{2}_{l}}{\textbf{k}^{2}_{||} + q^{2}_{l} + \omega_{n}^{2}}.
\end{equation}
The normal pressure is
\begin{equation}\label{2.08}
\text{P}_{\bot}=-\sigma_{33}=
\frac{1}{\beta L} \int\frac{d^{2}\textbf{k}_{||}}{(2\pi)^{2}} \;
\sum_{n,l \in Z} \;
\frac{q^{2}_{l}}{\textbf{k}^{2}_{||} + q^{2}_{l} + \omega_{n}^{2}}.
\end{equation}
{One can start summations over the index $n$ in (\ref{2.08}) instead of $l$ (as opposed to the way it was done in (\ref{sum over l})). Then using (\ref{1.84}) one gets the following
\begin{equation}
\text{P}_{\bot}=\frac{1}{2L}\sum_{l \in Z}\,\int\frac{d^{2}\textbf{k}_{||}}{(2\pi)^{2}} \; \frac{q_l^2\, \text{cth}\bigg(\frac{\beta}{2}\sqrt{k_{||}^2+q_l^2}\bigg)}{\sqrt{k_{||}^2+q_l^2}},  
\end{equation}
that can be rewritten as
\begin{equation}\label{2.07}
\text{P}_{\bot}=\frac{1}{2L}\sum_{l \in Z}\,\int\frac{d^{2}\textbf{k}_{||}}{(2\pi)^{2}} \; \frac{q_l^2\, \bigg(\text{cth}\bigg(\frac{\beta}{2}\sqrt{k_{||}^2+q_l^2}\bigg)-1\bigg)}{\sqrt{k_{||}^2+q_l^2}}+\lim_{T\rightarrow 0}\text{P}_{\bot},  
\end{equation}
where 
\begin{equation}
\lim_{T\rightarrow 0}\text{P}_{\bot}=\frac{1}{2L}\sum_{l \in Z}\,\int\frac{d^{2}\textbf{k}_{||}}{(2\pi)^{2}} \; \frac{q_l^2}{\sqrt{k_{||}^2+q_l^2}},    
\end{equation}
that is a divergent sum over $l$. To calculate this sum, it is necessary to introduce the cutoff function as it was done in (\ref{cutoff sum}). However, the normal pressure at zero temperature can be calculated using expressions (\ref{1.81}) and (\ref{casimir_force}). Then we have the following
\begin{equation}
\lim_{T\rightarrow 0}\text{P}_{\bot}=-\frac{\pi^2}{240L^4} - \frac{d \mathscr{E}_{vac}}{d V}.  
\end{equation}
The integral in the first term on the right hand side of the equation (\ref{2.07}) can be calculated analytically. Thus, one gets}
\begin{equation}\label{Pbot}
\begin{aligned}
\text{P}_{\bot} = -\frac{\pi^2}{240\, L^4}\; - \frac{d\mathscr{E}_{vac}}{dV}\;
-\;\frac{\pi T}{L^3} \sum_{l = 1}^{\infty} \; 
l^2\; \ln\bigg(1 - \exp \bigg(- \frac{\pi l}{LT}\bigg) \bigg)
\end{aligned}
\end{equation}
that can be rewritten as
\begin{equation}\label{Pbot_}
\begin{aligned}
\text{P}_{\bot} = -\frac{\pi^2}{240\, L^4}\; - \frac{d\mathscr{E}_{vac}}{dV}\;
+\;\frac{ 1}{\beta\pi L} \sum_{l = 1}^{\infty} \;  q_l^2\text{Li}_{1}(e^{-\beta q_{l}}),
\end{aligned}
\end{equation}
where $\text{Li}_1(x)=\sum_{k=1}^{\infty}x^{k}/k=-\ln(1-x)$. We can see that the expression (\ref{Pbot_}) is an expansion in a power series of $e^{-\frac{\pi}{LT}}$, and expression (\ref{2.06}) is also an expansion in a power series, but of $e^{-\pi LT}$. Thus, expressions (\ref{Pbot_}) and (\ref{2.06}) provide good descriptions of different asymptotic regimes. In particular, (\ref{Pbot_}) is suitable for small values of $L$ and $T$, where the contribution from the quantum vacuum dominates and thermal corrections are rather small. Conversely, (\ref{2.06}) applies to large values of $L$ and $T$, when the classical contribution (zero Matsubara mode term) dominates and quantum fluctuations are small.
\par
Also let us note that the first term in the right hand side of (\ref{Pbot}) is the standard Casimir force per unit area occurred at $T=0$, the second term is the ubiquitous "vacuum pressure", caused by vacuum energy density, which can be ignored without affecting the results, the third term is the normal pressure of the black body radiation in a slit-like pore. Eq. (\ref{Pbot}) yields the following bulk pressure (\ref{1.81}).
Thus, we can obtain the disjoining pressure as follows
\begin{equation}\label{Mehra}
\begin{aligned}
\Pi
= \text{P}_{\bot}-\text{P}_{\bot}^{(b)}=\Pi_{C} +\Pi_{B},
\end{aligned}
\end{equation}
where $\Pi_{C}$ is the already obtained Casimir disjoining pressure (\ref{casimir_force}) and
\begin{equation}\label{}
\begin{aligned}
\Pi_{B} = -\frac{\pi T}{L^3} \sum_{l = 1}^{\infty} \; 
l^2\; \ln\bigg(1 - \exp \bigg(- \frac{\pi l}{LT}\bigg) \bigg)-\frac{\pi^2 T^4}{45},
\end{aligned}
\end{equation}
is the contribution to the total disjoining pressure of the black body radiation.
Note that eq. (\ref{Mehra}) is the result obtained for the first time by Mehra~\cite{mehra1967temperature}. In SI units one can write this as follows
\begin{equation}\label{}
\begin{aligned}
\Pi_{C} = -\frac{\pi^2\hbar c}{240 L^4},
\end{aligned}
\end{equation}
\begin{equation}\label{}
\begin{aligned}
\Pi_{B} = -\frac{\pi k_B T}{L^3} \sum_{l = 1}^{\infty} \; 
l^2\; \ln\bigg(1 - \exp \bigg(- \frac{\pi\hbar c \,l}{k_BTL}\bigg) \bigg)-\frac{\pi^2 k_B^4T^4}{45\hbar^3 c^3}.
\end{aligned}
\end{equation}

\subsubsection{Tangential pressure}

Using (\ref{1.78}) and starting summations over $n$ in the expression (\ref{1.82}) one drives the following
\begin{equation}\label{P||}
\begin{aligned}
\text{P}_{||} = \frac{\pi^2}{720\, L^4}\; - \frac{d\mathscr{E}_{vac}}{dV} \;-\\
- \; \frac{1}{\pi\beta L} \sum_{l = 0}^{\infty}{}' \; \int\limits_{0}^{\infty}dk_{||} \; k_{||}  \;
 \ln\bigg(1 - \exp\bigg(-\beta\sqrt{k^{2}_{||} + q_l^2}\bigg)\bigg).
\end{aligned}
\end{equation}
Note that at $L\to \infty$
\begin{equation}
\text{P}_{||} =\text{P}_{||}^{(b)}=\text{P}_{\bot}^{(b)}=\frac{\pi^2 T^4}{45} - \frac{d\mathscr{E}_{vac}}{dV}
\end{equation}
that is in accordance with the fact that the photon gas in the thermodynamic limit is isotropic (see also Appendix \ref{Tangential pressure in the bulk}).

Expanding the logarithm in power series in the integrand in (\ref{P||}) in accordance with
\begin{equation}
\ln(1-x)=-\sum\limits_{k=1}^{\infty}\frac{x^k}{k}
\end{equation}
and taking the integral over $k_{||}$ in each term, we obtain
\begin{equation}\label{P_tang}
\begin{aligned}
\text{P}_{||} = \frac{\pi^2}{720\, L^4}\; - \frac{d\mathscr{E}_{vac}}{dV} \;+\frac{\zeta(3)}{2\pi\beta^3L}+\frac{1}{\pi\beta^3 L}\sum\limits_{l=1}^{\infty}\left(\text{Li}_{3}(e^{-\beta q_l})+\beta q_l \text{Li}_{2}(e^{-\beta q_l})\right).
\end{aligned}
\end{equation}
The first term on the right-hand side of equation (\ref{P_tang}) represents the tangential stress occurred at $T=0$, which is analogous to the Casimir disjoining pressure (\ref{casimir_force}). The second term is determined aforementioned vacuum pressure. The third and fourth terms represent thermal corrections to the Casimir tangential stress caused by the black body radiation. {Eq. (\ref{P_tang}) is a new fundamental result of this paper.}

Note that the tangential and normal pressures, which are determined by equations (\ref{Pbot}) and (\ref{P||}), can also be derived from standard thermodynamic relations 
\begin{equation}\label{sigma11}
\sigma_{11}=-P_{||}  =\frac{1}{LL_y}\frac{\partial F}{\partial L_x}  ,  \quad \sigma_{22}=-P_{||} =\frac{1}{LL_x}\frac{\partial F}{\partial L_y}  ,
\end{equation}
\begin{equation}\label{sigma33}
\sigma_{33}=-P_{\bot} =\frac{1}{L_xL_y}\frac{\partial F}{\partial L},
\end{equation}
where the free energy is determined by eqs. (\ref{1.98}-\ref{2.00}). This also indicates the consistency between the stress tensor approach and the purely thermodynamic approach. However, in general, when calculating the Casimir force between conducting bodies of arbitrary geometry, it becomes difficult to apply a purely thermodynamic method. In such cases, it is preferable to use the stress tensor, supplemented by a numerical solution of the equations for the propagators (\ref{1.67}) and (\ref{1.25}).

\section{Concluding remarks}
In conclusion, our paper introduces a quantum theory that explains Casimir forces between perfect electric conductors at finite temperatures. This theory is derived from the fundamental principles of quantum electrodynamics and quantum statistical physics. By applying Kapusta's finite-temperature quantum field theory and the Faddeev-Popov ghost formalism, we calculated the Casimir force at different temperatures. This allowed us to confirm existing findings and gain new insights into the theory. We have also derived the total stress tensor associated with Casimir forces, which is consistent with the free energy of the quantum electromagnetic field in thermodynamic equilibrium with the metal walls of the slit-like pore. Specifically, by using this tensor, we have calculated the tangential pressure on the surfaces of the conductive, slit-like pores due to the Casimir effect. 

The developed theoretical background can be used to calculate the interaction between conductive surfaces of different shapes. We believe that this could serve as a theoretical basis for estimating the Casimir forces that occur in various microdevices used in modern nanoengineering and nanotechnologies~\cite{chan2001quantum,palasantzas2020applications,elsaka2024casimir}.

\section*{Acknowledgments}
The authors would like to extend their sincere gratitude to Rudolf Podgornik for his valuable insights and motivating comments on our current findings. Regrettably, he passed away at the end of 2024, and this paper is dedicated in his memory. The authors also thank the anonymous reviewer for their valuable comments and suggestions, which have allowed us to improve the paper. The authors thank the Russian Science Foundation (Grant No. 24-11-00096) for financial support. 

\textbf{Data availability statement.} {\sl The data that supports the findings of this study are available within the article.}

\appendix

\section{Eigenfunctions}\label{Eigenfunctions}
It is easy to check that the functions $f$ and $u$ satisfy the completeness condition using the equalities
\begin{equation}\label{1.62}
\begin{aligned}
\frac{2}{L}\sum^{\infty}_{l=1}\sin\bigg(\frac{\pi l \xi}{L}\bigg)\sin\bigg(\frac{\pi l \xi'}{L}\bigg) = \delta(\xi-\xi'),
\end{aligned}
\end{equation}
\begin{equation}\label{1.63}
\begin{aligned}
\frac{2}{L}\sum^{\infty}_{l=0}{}'\cos\bigg(\frac{\pi l \xi}{L}\bigg)\cos\bigg(\frac{\pi l \xi'}{L}\bigg) = \delta(\xi-\xi'),
\end{aligned}
\end{equation}
\begin{equation}\label{1.64}
\begin{aligned}
\frac{1}{\beta}\sum_{n\in Z}e^{\frac{2\pi n i}{\beta}(\xi-\xi')}  = \delta(\xi-\xi'),
\end{aligned}
\end{equation}
\begin{equation}\label{}
\begin{aligned}
\frac{1}{2 \pi}\int\limits_{-\infty}^{+\infty} dk e^{ik(\xi-\xi')}  = \delta(\xi-\xi'),
\end{aligned}
\end{equation}
where $\xi,\xi'\in (0,L)$, the dashed sum means that the term corresponding to $l=0$ enters with an additional factor $1/2$, and thus $f$ and $u$ form an orthonormal basis, since
\begin{equation}\label{}
\begin{aligned}
\frac{2}{L}\int\limits_{0}^{L} d\xi\sin\bigg(\frac{\pi l \xi}{L}\bigg)\sin\bigg(\frac{\pi l' \xi}{L}\bigg) = \delta_{ll'},
\end{aligned}
\end{equation}
\begin{equation}\label{}
\begin{aligned}
\frac{2}{L}\int\limits_{0}^{L} d\xi \cos\bigg(\frac{\pi l \xi}{L}\bigg)\cos\bigg(\frac{\pi l' \xi}{L}\bigg) = \delta_{ll'},
\end{aligned}
\end{equation}
\begin{equation}\label{}
\begin{aligned}
\frac{1}{\beta}\int\limits_{0}^{\beta} d\xi e^{\frac{2\pi \xi i}{\beta}(n-n')}  = \delta_{nn'},
\end{aligned}
\end{equation}
\begin{equation}\label{}
\begin{aligned}
\int\limits_{-\infty}^{+\infty} \frac{d\xi}{2\pi} e^{i\xi(k-k')}  = \delta(k-k'),
\end{aligned}
\end{equation}
$l,l'=1,2,3\dots$, $n,n'\in Z\dots$. Therefore,
\begin{equation}\label{}
\begin{aligned}
\int\limits_{0}^{\beta} d\tau\int\limits_{V} d^{3}\textbf{x} \, f(x,\textbf{k}_{||},n,l)f(x,\textbf{k}'_{||},n',l') = \delta_{ll'}\delta_{nn'}\delta(k-k'),
\end{aligned}
\end{equation}
\begin{equation}\label{}
\begin{aligned}
\sum^{\infty}_{l=0}\sum_{n\in Z} \int d^{2}\textbf{k}_{||} \; f(x,\textbf{k}_{||},n,l)f(x',\textbf{k}'_{||},n',l') = \delta(x-x'),
\end{aligned}
\end{equation}
The functions $u(x,\textbf{k}_{||},n,l)$ satisfy similar conditions of completeness and normalization:
\begin{equation}\label{}
\begin{aligned}
\sum^{\infty}_{l=0}\sum_{n\in Z} \int d^{2}\textbf{k}_{||} \; u(x,\textbf{k}_{||},n,l)u(x',\textbf{k}'_{||},n',l') = \delta(x-x').
\end{aligned}
\end{equation}

\section{Tangential pressure in the bulk} \label{Tangential pressure in the bulk}

Using (\ref{1.90}) it can be shown that
\begin{equation}\label{}
\text{P}_{||}^{(b)} = T \,\sum_{n = 0}^{\infty}{}'\,\int\frac{d^{2}\textbf{k}_{||}}{(2\pi)^{2}} \;
\frac{k^{2}_{||}}{\sqrt{k^{2}_{||} + \omega_{n}^{2}}}.
\end{equation}
that using (\ref{identity1}) can be rewritten as 
\begin{equation}\label{}
\text{P}_{||}^{(b)} = 2T\lim_{L_x\rightarrow \infty}\left(\frac{1}{L_x} \sum_{l_x \in Z} \sum_{n = 0}^{\infty}{}'\,\int\limits_{-\infty}^{+\infty}\frac{dk_{2}}{2\pi} \;
\frac{q^2(l_x)}{\sqrt{k^{2}_{2} + \omega_{n}^{2} + q^2(l_x)}}\right),
\end{equation}
where
\begin{equation}
q(l_x) = \frac{2\pi l_x}{L_x}.    
\end{equation}
From this one gets 
\begin{equation}\label{}
\text{P}_{||}^{(b)} = -2T\lim_{L_x\rightarrow \infty}\left(\frac{\partial w}{\partial L_x}\right),
\end{equation}
where
\begin{equation}
w(L_x) = \sum_{l_x \in Z} \sum_{n = 0}^{\infty}{}'\,\int\limits_{-\infty}^{+\infty}\frac{dk_{2}}{2\pi} \; \sqrt{k^{2}_{2} + \omega_{n}^{2} + q^2(l_x)}.
\end{equation}
For large $L_x$ one can do the substitution 
\begin{equation}
\sum_{l_x \in Z} \rightarrow L_x\int\limits_{-\infty}^{+\infty}\frac{dk_{1}}{2\pi}    
\end{equation}
and from this it follows that one can put
\begin{equation}\label{}
w(L_x) =  L_x\sum_{n = 0}^{\infty}{}'\,\int\frac{d^{2}\textbf{k}_{||}}{(2\pi)^{2}} \;
\sqrt{k^{2}_{||} + \omega_{n}^{2}},
\end{equation}
and introducing the cutoff function one gets
\begin{equation}\label{}
\text{P}_{||}^{(b)} =-2T\;\sum_{n = 0}^{\infty}{}' \; v(n).
\end{equation}
where $v(n)$ is defined by (\ref{v(n)}) thus
\begin{equation}
\text{P}_{||}^{(b)}=\text{P}_{\bot}^{(b)},\quad \text{P}_{\bot}-\text{P}_{||}=\Pi-\Pi_{||}.  
\end{equation}

\selectlanguage{english}
\bibliography{name}

\end{document}